\documentclass[prc,preprint,aps]{revtex4}
\usepackage{graphicx}
\begin{document}

\title{Landau-Ginzburg method applied to   finite fermion systems: Pairing
in Nuclei}
\author{M. K. G. Kruse$^1$, H. G. Miller$^1$, A. R. Plastino$^1$ ,\\ A.  Plastino$^{1,\,2}$ 
and S. Fujita$^3$}
\affiliation{$^1$ 
    Department of Physics, University of Pretoria,Pretoria 0002,        South 
Africa\\$^2$ La Plata Physics Institute, National University La Plata and CONICET\\C. C. 727, 1900 La Plata, 
Argentina\\
$^3$ Department of Physics, SUNY at Buffalo, Buffalo, New York, USA}

\begin{abstract}
Given the  spectrum of a Hamiltonian,  a methodology is developed which employs the 
 Landau-Ginsburg method for characterizing  phase transitions  in infinite systems
  to  identify  phase transition remnants  
in finite fermion systems. As a first application of our appproach we discuss pairing in finite nuclei. \\
\noindent
\vspace{2mm}

\end{abstract}

\maketitle

Recently it has been pointed out that empirical evidence exists for a pairing 
phase transition to occur in symmetric nuclear matter at normal nuclear 
densities at $T_c \approx$8 Mev \cite{DHMMT93,RMK01}. Here the energy density 
and the specific heat have been obtained from a finte temperature extension of 
the semi-empirical mass formula\cite{S61}. A Landau Ginzburg treatment 
of this transition together with a simple pairing calculation
strongly suggests that its origin is due to the existence of
a paired superconducting phase at temperatures below .8 Mev\cite{LP80}. This 
result is not surprising as finite temperature BCS calculations in nuclear and 
neutron matter have suggested that such a phase should 
exist\cite{BCLL90,L80,G69,BPPR69,CC69,C69,O70,YC71,CCY72,T72,AO85,ES60}. \

One of the intriguing questions that arises is whether or not a remnant of this
phase transition survives in finite nuclei. Clearly a universal feature of
finite nuclei is a significant change in the density of states at excitation 
energies
of 10 MeV or less\cite{GC65}.  At lower excitation energies the spectrum of 
most nuclei is sparse and dominated by a relatively small number of collective 
states. With increasing excitation energy, the independent particle degrees of 
freedom dominate and the density of states grows exponentially. As the mass 
number increases, the low-lying collective portion of the energy
spectrum becomes more compressed and an abrupt change in the many particle
density of states occurs at lower excitation energies. It has, therefore, been 
suggested
that a collective to non-collective phase transition occurs in finite
nuclei~\cite{MCQ89}. Strictly speaking it is incorrect to speak of phase
transitions in finite systems. No one can deny, however, that transitions between
different regimes do take place in these systems, and that they are more or less
abrupt. (A similar situation is known in biophysics. The helix-coil transition in certain biological 
molecules in solution occurs with the temperature width of around 5 degrees.)
 This issue, particularly in deformed systems, has been clouded by the fact that
finite temperature mean field calculations have suggested that this phase
transition is simply  due to a drastic change of shape \cite{LA84}.  The
deformed-to-spherical shape transition seen in these calculations is not seen in
exact canonical calculations~\cite{MQC89,DPR91,RQM92,CS94} and may be an
artifact of the finite temperature mean field approximation and also depends  on
the volume of the system~\cite{YM92,YM94}.
In spite of the fact that  the canonical partition function above the
critical  temperature is dominated by the single particle degrees of freedom a
few  collective states still contribute and are extremely important in
 calculation of shape dependent parameters. Recent calculations of the ensemble
average of the
quadrupole moment squared $ Q^{[2]}\cdot $ $Q^{[2]}$ indicate that it
is discontinuous in the finite temperature mean field approximation,
while no discontinuity is observed in the canonical
calculations~\cite{CMQ94}. In both cases this quantity does not appear to vanish
at the critical temperature. It should also be noted, however, that when thermal
fluctuations
in the shape dependent order parameters are taken into account, either by
macroscopic or microscopic procedures, reasonable agreement with the exact
canonical calculations~\cite{MQC89} is obtained. With increasing temperature,
however, it is expected that these collective degrees of freedom will eventually
completely dissolve, presumably below the critical
temperature of the liquid-to-gas phase transition.

However, it should be noted that model studies in a SU(2)$\times$SU(2) system show that,
in the thermodynamic limit, this system exhibits a singularity in the specific heat 
characteristic of a true phase transition\cite{FGN79,RP85}. Furthermore, the remnant of this 
singularity remains in the form of a peak in finite sytems of this type. The presence of this 
peak has been used to map out the phase structure in such a model\cite{DM87}.

In spite of the fact that in many microscopic variational calculations (see for example \cite{DB04})
pairing transitions appear to take place they are difficult to identify in exact shell model calculations \cite{DH03}.
We propose in the present work to identify the existence of a pairing
transition empirically  in finite nuclei in the following manner. In the 
Canonical
Ensemble the partition function for a nucleus of mass A is given by 
\begin{equation}
{\sf Z} (A,T) = \sum^n_i {g_i \ exp(-\beta E_i)}
+\int^{E_{max}}_{E_n} dE \
g_{A,Z}(E) \  exp(-\beta E) \end{equation}

\noindent  where
$\beta=\frac{1}{T}$, \
$g_i=2j_i + 1$ is the spin degeneracy factor, $E_i$ the  energy of the $i$th 
state of the nucleus and
$g_{A,Z}(E)$ its level density with 
\begin{equation}
\label{ge}
 {g_{A,Z} (E)= \frac{\sqrt{\pi}}{12} \frac{exp(2\sqrt{aU})}
{a^{\frac{1}{4}}}U^{\frac{5}{4}}} 
\end{equation}
\noindent
where  $U = E - P(Z) - P(N)$ and $a = A[0.00917S + 0.142]$  per MeV for
undeformed nuclei, or $a = A[0.00917S + 0.120]$ per MeV for deformed nuclei. 
$A$ is the mass number
and $S$ is the spin energy.
Given experiment information about the bound 
states of a nucleus and an experimental fit to its continuum level density, the 
partition function of a nucleus can be determined empirically

From $\sf Z$ it is easy to determine as a function of T the excitation energy 
\begin{equation}{\sf E}=-\frac{\partial}{\partial \beta} ln{{\sf 
Z}}\end{equation}
\noindent
and the specific heat

\begin{equation}
\label{sh}
{\sf C_V}=\frac{\partial {\sf E}} {\partial T}.\end{equation}

\noindent
A second order phase transition such as a pairing phase transition in an 
infinite system should lead to a discontinuity in ${\sf C}_V$ at $T_c$. Clearly 
such behavior is not posible in finite sytems. Only a remnant remains in the 
form of peaked structure in the specific heat.  In order to ascertain whether 
this structure is consistent with a pairing phase transition we make use of 
Landau-Ginzburg theory to demonstrate that  a description of the condensed or 
paired phase close to the critical temperature, $T_c$, from information in the 
normal or uncondensed phase obtained from experimental data can be can be 
accomplished.  The magnitude of the remnant of this discontinuity in ${\sf
C}_V$ is compared with a simple analytical calculation for a pairing phase 
transition\cite{LP80}  corrected for finite size effects\cite{P84}. \

Landau and Ginzburg have provided a simple theory of phase transitions which 
approximates the free energy in the region around $T_{c }$ and is most useful 
in analyzing
the thermodynamics in this region. In particular, using only knowledge about the 
uncondensed phase one is able to make predictions about quantities in the 
condensed phase,such as specific heat, magnetic susceptibility and 
compressability. Moreover, Landau-Ginzburg theory can be derived from 
microscopic considerations \cite{LP80}. \

In the Landau Ginzburg formulation it is necessary first to determine an 
expression for the free energy  ${\sf F(T)}$ in both phases.  In the following 
the subscript 1 will refer to the lower temperature (condensed) phase, and 2 to 
the higher temperature (uncondensed or normal) phase. In the uncondensed phase, 
a quadratic form for the  energy,
which follows from a low temperature Fermi gas, is used approximation of a 
normal
Fermi liquid, \begin{equation}{\sf E}_2(T)=a_2 + k_2 T^2, 
\label{eqE2}\end{equation}where $a_2$ and $k_2$ are constants. From the 
relations for the specific heat  terms of ${\sf E}$ and the entropy  ${\sf 
S}$,\begin{equation}{\sf C}_V=\frac{\partial {\sf E}}{\partial T} = T\, 
\frac{\partial {\sf S}}{\partial T},\end{equation}
one obtains the entropy  in the uncondensed phase,
\begin{equation}
{\sf S}_2(T)=C_2 + 2 k_2 T, \label{eqs2}
\end{equation}
where $C_2$ is an unknown integration constant which later cancels out
of the calculation. From eqs.~(\ref{eqE2})
and (\ref{eqs2}) the free energy in the uncondensed
phase is given by
\begin{equation}
{\sf F}_2(T)=a_2 - C_2 T - k_2 T^2. \label{eqf2}
\end{equation}
 The free energy  in the condensed phase is obtained from
 the Landau expansion  \cite{LP80} for the free energy
 in terms of an order parameter $\eta$ which goes to zero at 
the transition to the uncondensed phase. This order parameter vanishes at a 
critical temperature
$T_c$.  The free energy expansion to order $\eta^4$ is
\begin{equation}
{\sf F}_1(T,\eta)={\sf F}_2 + A \eta^2 + B \eta^4.
\end{equation}
Here $A$ and $B$ are functions of temperature and it is assumed that the states 
with $\eta$=0 and $\eta \neq $0 are of different symmetry.
In this case it can be shown
the linear term in $\eta$ must be set equal to zero and
if the critical point is also a stable point, e.g. if ${\sf F}_1$ as a function 
of $\eta$ is a minimum at $\eta$ =0, then the third order term in $\eta$ should 
be zero and at the critical point \cite{LP80}
\[ A=0 \hspace{10mm} B > 0
\]
The order parameter is determined by requiring the condensed phase
to be stable below $T_c$ (i.e.\ ${\sf F}_1$ should be minimized
w.r.t.\ $\eta$).  This leads to
\begin{equation}
{\sf F}_1={\sf F}_2 -\frac{A^2}{4 B}.
\end{equation}
Furthermore, since
$A$ is of opposite sign in the condensed and uncondensed phases,
while $B$ is strictly positive \cite{LP80},
the lowest order expansion of $A$ in $T-T_c$ can be parametrized as
\begin{equation}
A(T)= a (T-T_c)\; 2 \sqrt{B(T_c)}.
\end{equation}
Note especially that $a>0$ is an essential requirement following from
the phase diagram \cite{LP80}.
Substituting for $A(T)$, the free energy per nucleon near $T_c$ is
given by
\begin{equation}
{\sf F}_1(T)= (a_2-a^2 T_c^2) + (2a^2T_c-C_2)T -(k_2 + a^2) T^2, \label{eqf1}
\end{equation}
where ${\sf F}_2$ is taken from eq.~(\ref{eqf2}).

From  eq.~(\ref{eqf1}),
 the energy  in the condensed is easily determined to be phase near $T_c$,
\begin{eqnarray}
{\sf E}_1(T)&=& (a_2 - a^2 T_c^2) + (a^2 + k_2) T^2\\
                &=&  a_1 + k_1 T^2 .\label{eqW1}
\end{eqnarray}
Comparing this to the uncondensed phase (eq.~(\ref{eqE2})) we
note that the $T$ dependence is also quadratic, but has a larger
coefficient.  Thus the specific heat is discontinuous across the
phase transition, and is necessarily larger ($k_1 > k_2$)  in the
condensed phase.     

We now apply our theoretical results to the following even-even nuclei, namely, 
$^{20}$Ne, $^{48}$Ca, $^{88}$Sr and $^{208}$Pb. These  nuclei span a
large portion  of the mass  spectrum and  have a reasonably well determined 
energy spectrum both in terms of their energies and the corresponding angular 
momentum assignments.
For each of these nuclei,
the partition function (see eq. (1)) contains a discrete  as well as a 
continuum contribution. 

 The discrete portion of the partition function is determined empirically
from the measured energy levels in each nuclei\cite{GC65}. At lower excitation 
energies the angular momemtum assignment to each state in the spectrum is 
generally unique. As the energy rises, predominantly   
 near the onset of the  continuum, states occur which have an 
uncertainty in the angular momentum assignment.  In such  cases, we have taken  
the lowest suggested value of the angular momentum. The continuum contribution 
to the partition function is given by equation (\ref{ge}) and the value of the 
parameters used for the even-even-nuclei considered are given in Table 1.

 The energy at which the continuum is to be attached to the discrete portion of 
the energy spectrum is determined via the prescription of Gilbert and 
Cameron\cite{GC65}. This 
matching point was ascertained graphically by first  plotting the energy as a 
function of the number of levels. There are two parts in each such graph, viz, 
the curve corresponding to the low-energy bound states, and the curve 
corresponding to the high energy continuum states. The point of tangency (i.e. 
where the slope is the same in both curves) was then determined and selected as 
the matching point. They found that this  could be parameterized by the 
following\begin{equation}
E_{c}=U_{x}+E_{p}
\end{equation}
In the above equation, $E_{c}$ is the energy at the matching point 
and $E_{p}$ is the pairing energy associated with the nucleus under 
consideration. $U_{x}$ is an additional energy term which is found graphically 
from the tangency point (minus the pairing energy) as a function of mass number 
(see Gilbert and Cameron\cite{GC65}). From the graphical results one can see 
that there is an upper and lower limit to $U_{x}$, that the curves are 
hyperbolas, and that the overall behaviour is a decrease in $U_{x}$ with 
increasing $A$. The range for $U_{x}$ is listed below. \
On average, 

$U_{x}=2.5 + \frac{150}{A}$, \

the upper limit is , $U_{x}=2.7 + \frac{200}{A}$, \

and the lower limit is, $U_{x}=2.1 + \frac{120}{A}$. \\
\noindent
We  have used the upper limit in all our calculations, except for $^{208}$Pb as 
it produced the best fit to a straight line in the uncondensed portion of the 
specific heat (see figures 1-4).

For $^{208}$Pb it is extremely difficult to determine the density of states. It 
has been pointed out by Gilbert and Cameron that not all nuclei near closed 
shells could be fitted (i.e. finding the tangency point) in the aforementioned 
manner. For those that could be fitted, a value for $U_{x}$ was found that was 
much larger than the one predicted. $^{208}$Pb is such a case. One can determine 
$U_{x}$ from the graphical results given in Gilbert and Cameron (this value 
is roughly 4.8 MeV). Adding the pairing energy to this yields a matching point  
value of roughly 6.5 MeV. We used a slightly lower value. The parameters used in 
the determination of the partition function as a function of temperature for the 
nuclei under consideration are given in Table 1.

For all four nuclei (see Figures 1-4), a well defined peak is observed in the 
specific heat ( equation(\ref{sh})) as a function of temperature and the 
Landau-Ginzburg fits to the specific heat are consistent with that of a remnant 
of a second order phase transition.  The discontinuity in the specific heat 
expected in an infinite system appears to be smoothed over at the critical 
temperature due to finite size effects. Furthermore, not unexpectedly at higher 
tempertures above the critical temperature, the specifc heat is linear as a
function of temperature.

\begin{table}\caption{Parameters used for nuclei}
	\label{T:dimens}
	\begin{center}
        \begin{tabular}{|c|c|c|c|c|c|c|c|}
            \hline
            Nuclei & $P(Z)$ (MeV) & $P(N)$ (MeV) & $S(Z)$ (MeV) & $S(N)$ (MeV) 
& $a$ (MeV$^{-1}$) & $E_{c}$ (MeV) \\
            \hline
            $^{208}$Pb & 0.83 & 0.38 & -8.86 & -3.16 & 6.61 & 6.25  \\
	    $^{88}$Sr & 1.24 & 0.93 & -16.41 & 12.88 & 9.65 & 7.14 \\
	    $^{48}$Ca & 1.83 & 1.3 & -12.07 & 12.13 & 6.80 & 10 \\
	    $^{20}$Ne & 2.5 & 2.5 & -0.811 & 5.633 & 4.58 & 17.7 \\
            \hline
        \end{tabular}
	\end{center}
\end{table}

\begin{figure}
	\centering
	\includegraphics{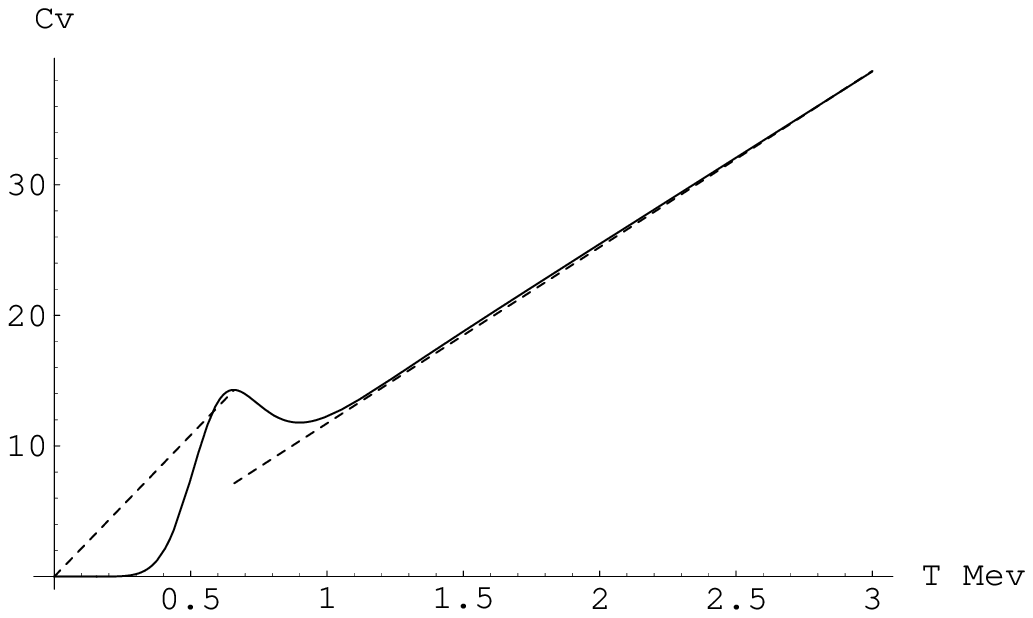}
	\caption{Specific heat of $^{208}$Pb as a function of temperature for the 
parameters given in Table 1. The dashed curve is the Landau-Ginzburg fit to the 
specific heat, whereas the solid curve is the specific heat determined from the 
experimental data. . $k1 = 10.83$ and $k2 = 6.75$}
\end{figure}

\begin{figure}
	\centering
	\includegraphics{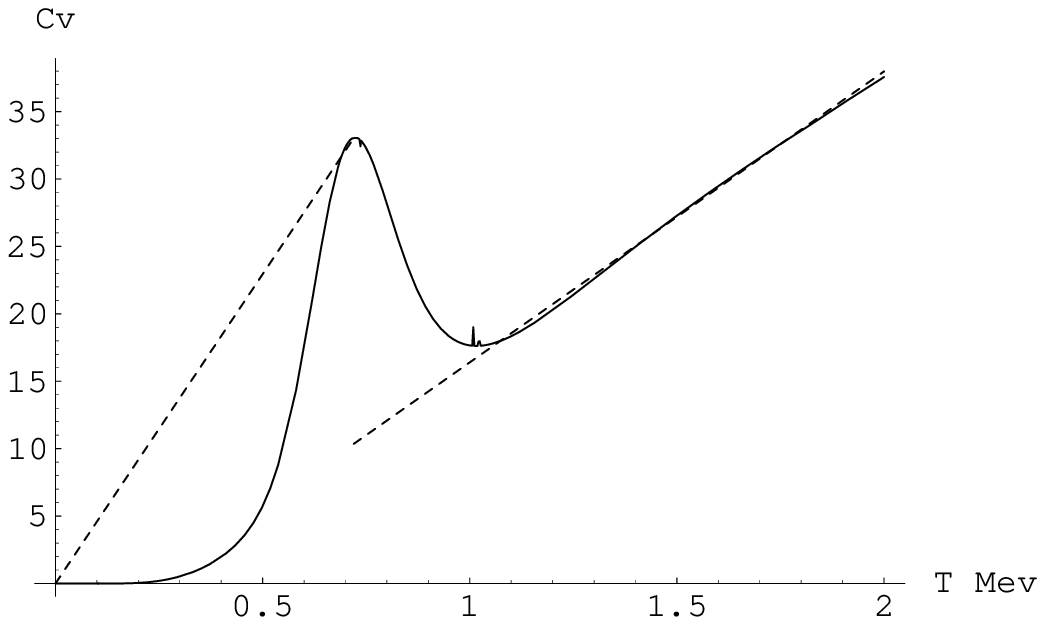}
	\caption{Specific heat of $^{88}$Sr as a function of temperature for the 
parameters given in Table 1. The dashed curve is the Landau-Ginzburg fit to the
specific heat, whereas the solid curve is the specific heat determined from the
experimetnal data. . $k1 = 22.94$ and $k2 = 10.79$}
\end{figure}

\begin{figure}
	\centering
	\includegraphics{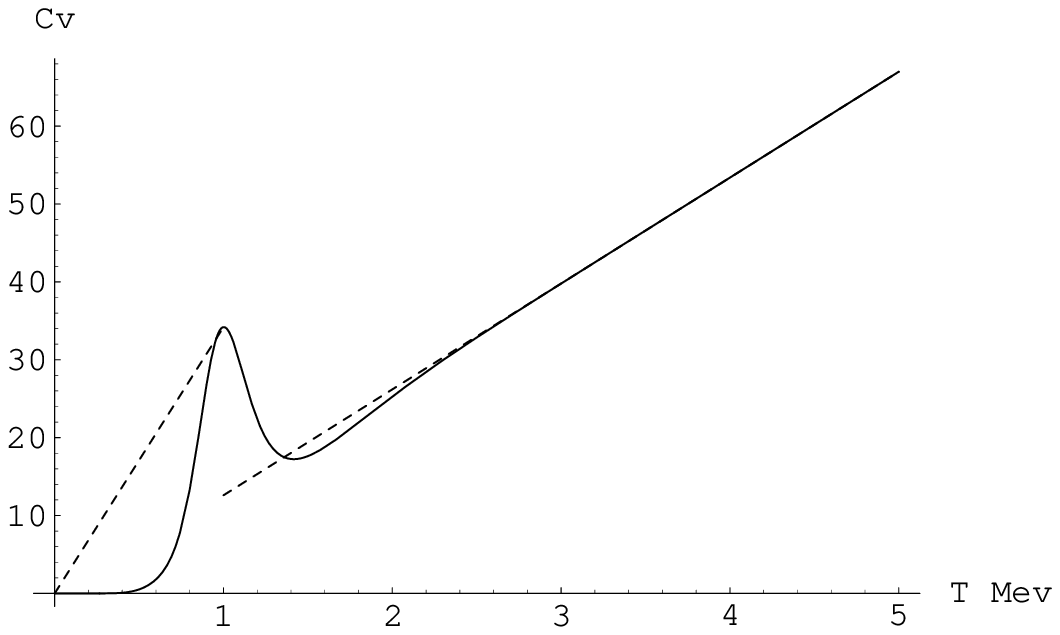}
	\caption{Specific heat of $^{48}$Ca as a function of temperature for the 
parameters given in Table 1. The dashed curve is the Landau-Ginzburg fit to the 
specific heat, whereas the solid curve is the specific heat determined from the 
experimental data. . $k1 = 17.10$ and $k2 = 6.80$}
\end{figure}

\begin{figure}
	\centering
	\includegraphics{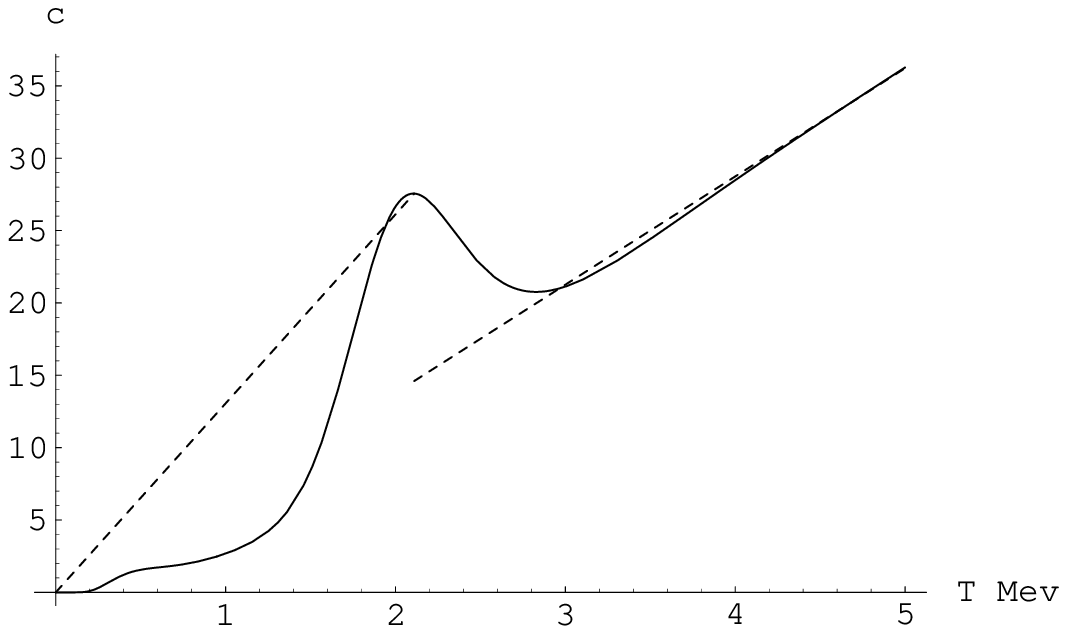}
	\caption{Specific heat of $^{20}$Ne as a function of temperature for the 
parameters given in Table 1. The dashed curve is the Landau-Ginzburg fit to the 
specific heat, whereas the solid curve is the specific heat determined from the 
experimental data. . $k1 = 6.53$ and $k2 = 3.74$}
\end{figure}

Varying the values of the matching points  for the 
four
nuclei considered gives rise no significant change in quantities 
like the critical temperature (variation of 0.01 MeV), or to the slope of the 
straight line fit (in the uncondensed phase) to the specific heat at higher 
temperatures. The numerical value of the observed peak  in the specific heat for 
each nucleus does vary slightly.

In an infinite system for
  a pairing phase transition, the discontinuity in the 
specific heat at $T_c$ can easily be analytically determined\cite{LP80} and is 
given by
\begin{eqnarray}
\frac{{\sf C}_s(T_c)-{\sf C}_n(T_c)}{{\sf C}_n(T_c)}
   &=&\frac{V\frac{4mp_fT_c}{7\zeta(3)\hbar^3}}{V\frac{mp_f T_c}{3\hbar^3}} \\     
    &=& \frac{12}{7\zeta(3)}\\    
     &=&1.43 
\end{eqnarray}
and one can define
\begin{eqnarray}
\Delta_L &=& {\sf C}_s(T_c)-{\sf C}_n(T_c)\\
  &=&1.43{\sf C}_n(T_c) 
  \label{dell}     
\end{eqnarray}
\noindent
Clearly in order to calculate the remnant of this discontinuity in ${\sf C}$ for
a nucleus, this result should be corrected for finite size effects. A simple 
way of modeling this is to assume that the equilibrated system is contained in 
a finite sized volume. The sum over the momentum states in quantities like the 
partition function can be approximated then by an integral over the density of 
states in the form ~\cite{P84,SMP91}
\begin{equation}
\Sigma_{p } \to \int\limits_{ }^{ } \frac{V}{2\pi^{2 } \hbar^3} p^2 d p
\pm \frac{S}{8\pi \hbar^2}pdp + \frac{L}{8\pi \hbar} dp\;,
\end{equation}
where $S$ and $L$ are the surface area and linear dimension of the system
of volume $V$. The $\pm$ correspond to either choosing Dirchlet (-)
or von Neumann (+) boundary conditions. In the following we shall use Dirchlet 
boundary conditions and neglect the contribution from the linear term.
 
 From Liftshitz and Pitaevskii \cite{LP80} it easy to see that there is no 
change in the numerator of equations (16 and 17) when finite size effects are 
included and that only the surface term must be taken into account in the 
calculation of ${\sf C}_n$ in the denominator. The surface contribution to
${\sf C}_n $ is given by   
\begin{equation}{\sf C}^{\it surf}_n = 
\frac{\partial {\sf E}^{\it surf}_n}{\partial T} 
\end{equation}
 where
 \begin{eqnarray}
 {\sf E}^{\it surf}_n &=& -\frac{2 S}{8 \pi \hbar^2}\int \frac{ \frac{p^2}{2m}
p dp}{z^{-1} \exp^\frac{\beta p^2}{2m} +1}\\ 
&=&-\frac{2 S m T^2}{8 \pi \hbar^2}\int 
\frac{x dx}{z^{-1}\exp^x+1}\\ 
&=& -\frac{2 S m T^2}{8 \pi \hbar^2} F_2(z)\\
&=& -\frac{2 S m T^2}{8 \pi \hbar^2}\frac{\epsilon^2_f}{2T^2}[1+\frac{2 \pi^2}{6 
\frac{\epsilon^2_f}{T^2}} \ldots]\ 
\end{eqnarray}
 \noindent
 where $\epsilon_f$ is the Fermi energy $z=e^{\beta \mu}$ is the fugacity and 
$F_2(z)$ is the well known Fermi-Dirac integral\cite{P84}. From this it follows 
that the surface contribution to the specific heat for finite nuclei is given 
by
\begin{equation}
{\sf C}^{\it surf}_v\approx -\frac{V m p_f T}{3 \hbar^3}[\frac{S \pi \hbar}{4 V 
p_f} \ldots]\end{equation}
and that 
\begin{eqnarray}
\frac{{\sf C}_s^{\it finite}(T_c)-{\sf C}_n^{\it finite}(T_c)}{{\sf   C}_n^{\it finite}(T_c)}
     &\approx&\frac{V\frac{4mp_fT_c}{7\zeta(3)\hbar^3}}{V\frac{mp_f T_c}{3\hbar^3} [1-\frac{3 S \hbar}{2 V p_f}]} \\     
&=&\frac{1.43}{[1-\frac{S \pi \hbar}{4 V p_f}]}
\end{eqnarray} 
and in same manner one can define
\begin{eqnarray}
\Delta^{finite}_L&=&{\sf C}_s^{\it finite}(T_c)-{\sf C}_n^{\it finite}(T_c)\\                
 &=&\frac{1.43}{[1-\frac{S \pi \hbar}{4 V p_f}]} {\sf   C}_n^{\it finite}(T_c)
\label{delfin}
\end{eqnarray}
\noindent
Here the the Fermi momentum, $p_F$, for a nucleus with $A$ nucleons
is determined from
\begin{equation}
A=\frac{8\pi p_f^3}{3(2\pi \hbar)^3}
\end{equation} \noindent
 and for simplicity $V=\frac{4\pi R^3}{3}$ and $S=4 \pi R^2$ where $R=r_0 
A^{\frac{1}{3}}$.

After the Landau-Ginburg fits to the  empirical values of the specific heat have 
been obtained (see figures 1-4) it is easy to determine
the  value of the remnant of the discontinuity (hereafter we shall refer to this 
as the discontinuity) in the  specific heat at $T_c$ for each of the nuclei (see 
TableII). As one goes to heavier nuclei $T_c$ falls off hyperbolically and reaches for $^{208}Pb$ 
a value
which is slightly less than that in symmetric nuclear matter\cite{RMK01}.
If one now assumes that this discontinuity is due to the remnant of a  
pairing phase transition, its value with($\Delta_L^{finite}$) and without finite 
size corrections ($\Delta_L$) can easily be obtained from  from equations 
(\ref{delfin}) and (\ref{dell}). In spite of the simplicity of the pairing 
calculation(the density of states and the pairing interaction are assumed to be 
constant)   and the difficulties with the continuum contribution (for $^{208}Pb$) 
the value of the  discontinuity calculated with or without finite 
size corrections is in reasonable agreement with the empirical value for the 
three heavier nuclei. Note that in the heavier nuclei the finite size corrections are not large.
 The discrepancy in all cases is less than $\approx 40\%$. 
Only in the case of $^{20}Ne$ is the discrepancy about 60\%,  more than 1.5 the 
value for heavier nuclei.  This strongly suggests the existence of a pairing 
transition in the heavier nuclei which is not present in $^{20}Ne$.  In the latter case 
this evidence for a second order transition which probably is shape related\cite{LA84}.

\begin{table}\caption{ The values of the  discontinuity in the specific heat 
at$T_c$}		\label{T:dimens}
	\begin{center}
        \begin{tabular}{|c|c|c|c|c|}
            \hline
            Nuclei & $\Delta$ & $\Delta_{L}$ & $\Delta_{L}^{finite}$ & $T_{c}$ 
(MeV) \\	    \hline
            $^{208}$Pb & 7.15 & 10.22 & 10.27 & 0.66 \\
	    $^{88}$Sr & 22.67 & 14.81 & 14.92 & 0.72 \\
	    $^{48}$Ca & 21.59 & 18.02 & 18.27 & 1.0 \\
	    $^{20}$Ne & 12.95 & 20.88 & 21.46 & 2.11 \\
            \hline
        \end{tabular}
	\end{center}
\end{table}

From the spectra of four even-even nuclei $^{20}Ne, ^{48}Ca, ^{88}Sr$ and $^{208}Pb$ we have empirically constructed 
in the Canonical Ensemble their partiton functions from available experimental data and determined their specific
heat as a function of temperature .  For each nuclei, the specific heat displays a prominent peak
which may be the remnant of a phase transition.  A Landau Ginzburg treatment  shows unambiguously that  this interpretation
is not inconsistent if the the phase transition is second order. A simple pairing calculation of the magnitude of the 
observed discontinuity
is consistent with that obtained empirically for the three heavier nuclei. This suggests that a pairing transition 
takes place in these 
nuclei.  In the case of $^{20}Ne$ such is not the case and the transition may be shape related.

Lastly we wish to point out that the methodology that we have employed here can be used for any finite fermion system. 
Given the  exact spectrum of the system,  the 
 Landau-Ginsburg method can be utilized to identify the existence of 
 the remnant of  a phase transition. In cases such as pairing phase transitions, simple analytical calculations 
can be used to identify the nature of the phase transition.

\vspace{8mm}
AP acknowledges support from the Argentine National Science Council.  HGM acknowledges the hospitality of 
the Physics Department of SUNY at Buffalo where part of this work was undertaken.

\newpage


\begin{thebibliography}{33}
\expandafter\ifx\csname natexlab\endcsname\relax\def\natexlab#1{#1}\fi
\expandafter\ifx\csname bibnamefont\endcsname\relax
  \def\bibnamefont#1{#1}\fi
\expandafter\ifx\csname bibfnamefont\endcsname\relax
  \def\bibfnamefont#1{#1}\fi
\expandafter\ifx\csname citenamefont\endcsname\relax
  \def\citenamefont#1{#1}\fi
\expandafter\ifx\csname url\endcsname\relax
  \def\url#1{\texttt{#1}}\fi
\expandafter\ifx\csname urlprefix\endcsname\relax\def\urlprefix{URL }\fi
\providecommand{\bibinfo}[2]{#2}
\providecommand{\eprint}[2][]{\url{#2}}

\bibitem[{\citenamefont{Davidson et~al.}(1993)\citenamefont{Davidson, Hsaio,
  Markram, Miller, and Tzeng}}]{DHMMT93}
\bibinfo{author}{\bibfnamefont{N.~J.} \bibnamefont{Davidson}},
  \bibinfo{author}{\bibfnamefont{S.~S.} \bibnamefont{Hsaio}},
  \bibinfo{author}{\bibfnamefont{J.}~\bibnamefont{Markram}},
  \bibinfo{author}{\bibfnamefont{H.~G.} \bibnamefont{Miller}},
  \bibnamefont{and} \bibinfo{author}{\bibfnamefont{Y.}~\bibnamefont{Tzeng}},
  \bibinfo{journal}{Phys.\ Lett.\ B} \textbf{\bibinfo{volume}{315}},
  \bibinfo{pages}{12} (\bibinfo{year}{1993}).

\bibitem[{\citenamefont{Ritchie et~al.}(2001)\citenamefont{Ritchie, Miller, and
  Khanna}}]{RMK01}
\bibinfo{author}{\bibfnamefont{R.~A.} \bibnamefont{Ritchie}},
  \bibinfo{author}{\bibfnamefont{H.~G.} \bibnamefont{Miller}},
  \bibnamefont{and} \bibinfo{author}{\bibfnamefont{F.~C.}
  \bibnamefont{Khanna}}, \bibinfo{journal}{Eur. Phys. J.}
  \textbf{\bibinfo{volume}{A10}}, \bibinfo{pages}{97} (\bibinfo{year}{2001}).

\bibitem[{\citenamefont{Seeger}(1961)}]{S61}
\bibinfo{author}{\bibfnamefont{P.~A.} \bibnamefont{Seeger}},
  \bibinfo{journal}{Nucl. Phys.} \textbf{\bibinfo{volume}{25}},
  \bibinfo{pages}{1} (\bibinfo{year}{1961}).

\bibitem[{\citenamefont{Liftshitz and Pitaeevskii}(1980)}]{LP80}
\bibinfo{author}{\bibfnamefont{E.}~\bibnamefont{Liftshitz}} \bibnamefont{and}
  \bibinfo{author}{\bibfnamefont{L.~P.} \bibnamefont{Pitaeevskii}},
  \emph{\bibinfo{title}{Statistical Physics II}}
  (\bibinfo{publisher}{Pergamon}, \bibinfo{address}{Oxford},
  \bibinfo{year}{1980}).

\bibitem[{\citenamefont{Baldo et~al.}(1990)\citenamefont{Baldo, Cugnon,
  Lejeune, and Lombardo}}]{BCLL90}
\bibinfo{author}{\bibfnamefont{M.}~\bibnamefont{Baldo}},
  \bibinfo{author}{\bibfnamefont{J.}~\bibnamefont{Cugnon}},
  \bibinfo{author}{\bibfnamefont{A.}~\bibnamefont{Lejeune}}, \bibnamefont{and}
  \bibinfo{author}{\bibfnamefont{U.}~\bibnamefont{Lombardo}},
  \bibinfo{journal}{Nucl.\ Phys.\ A} \textbf{\bibinfo{volume}{515}},
  \bibinfo{pages}{409} (\bibinfo{year}{1990}).

\bibitem[{\citenamefont{Lacombe{\it \ et al.}}(1980)}]{L80}
\bibinfo{author}{\bibfnamefont{M.}~\bibnamefont{Lacombe{\it \ et al.}}},
  \bibinfo{journal}{Phys. Rev.} \textbf{\bibinfo{volume}{C21}},
  \bibinfo{pages}{861} (\bibinfo{year}{1980}).

\bibitem[{\citenamefont{Ginzburg}(1969)}]{G69}
\bibinfo{author}{\bibfnamefont{V.~L.} \bibnamefont{Ginzburg}},
  \bibinfo{journal}{J. Stat. Phys} \textbf{\bibinfo{volume}{1}},
  \bibinfo{pages}{3} (\bibinfo{year}{1969}).

\bibitem[{\citenamefont{Baym et~al.}(1969)\citenamefont{Baym, Pethick, Pines,
  and Ruderman}}]{BPPR69}
\bibinfo{author}{\bibfnamefont{G.}~\bibnamefont{Baym}},
  \bibinfo{author}{\bibfnamefont{C.}~\bibnamefont{Pethick}},
  \bibinfo{author}{\bibfnamefont{D.}~\bibnamefont{Pines}}, \bibnamefont{and}
  \bibinfo{author}{\bibfnamefont{M.}~\bibnamefont{Ruderman}},
  \bibinfo{journal}{Nature} \textbf{\bibinfo{volume}{224}},
  \bibinfo{pages}{872} (\bibinfo{year}{1969}).

\bibitem[{\citenamefont{Clark and Chao}(1969)}]{CC69}
\bibinfo{author}{\bibfnamefont{J.~W.} \bibnamefont{Clark}} \bibnamefont{and}
  \bibinfo{author}{\bibfnamefont{N.~C.} \bibnamefont{Chao}},
  \bibinfo{journal}{Lett. Nuovo Cim.} \textbf{\bibinfo{volume}{2}},
  \bibinfo{pages}{185} (\bibinfo{year}{1969}).

\bibitem[{\citenamefont{Clark}(1969)}]{C69}
\bibinfo{author}{\bibfnamefont{J.~W.} \bibnamefont{Clark}},
  \bibinfo{journal}{Phys. Rev. Lett.} \textbf{\bibinfo{volume}{23}},
  \bibinfo{pages}{1463} (\bibinfo{year}{1969}).

\bibitem[{\citenamefont{Ostgaard}(1970)}]{O70}
\bibinfo{author}{\bibfnamefont{E.}~\bibnamefont{Ostgaard}},
  \bibinfo{journal}{Nucl. Phys.} \textbf{\bibinfo{volume}{A154}},
  \bibinfo{pages}{202} (\bibinfo{year}{1970}).

\bibitem[{\citenamefont{Yang and Clark}(1971)}]{YC71}
\bibinfo{author}{\bibfnamefont{C.~H.} \bibnamefont{Yang}} \bibnamefont{and}
  \bibinfo{author}{\bibfnamefont{J.~W.} \bibnamefont{Clark}},
  \bibinfo{journal}{Nucl. Phys.} \textbf{\bibinfo{volume}{A174}},
  \bibinfo{pages}{49} (\bibinfo{year}{1971}).

\bibitem[{\citenamefont{Chao et~al.}(1972)\citenamefont{Chao, Clark, and
  Yang}}]{CCY72}
\bibinfo{author}{\bibfnamefont{N.}~\bibnamefont{Chao}},
  \bibinfo{author}{\bibfnamefont{J.~W.} \bibnamefont{Clark}}, \bibnamefont{and}
  \bibinfo{author}{\bibfnamefont{C.~H.} \bibnamefont{Yang}},
  \bibinfo{journal}{Nucl. Phys.} \textbf{\bibinfo{volume}{A179}},
  \bibinfo{pages}{320} (\bibinfo{year}{1972}).

\bibitem[{\citenamefont{Tatsuka}(1972)}]{T72}
\bibinfo{author}{\bibfnamefont{T.}~\bibnamefont{Tatsuka}},
  \bibinfo{journal}{Prog. Theor. Phys} \textbf{\bibinfo{volume}{48}},
  \bibinfo{pages}{1517} (\bibinfo{year}{1972}).

\bibitem[{\citenamefont{Amundsen and Ostgaard}(1985)}]{AO85}
\bibinfo{author}{\bibfnamefont{L.}~\bibnamefont{Amundsen}} \bibnamefont{and}
  \bibinfo{author}{\bibfnamefont{E.}~\bibnamefont{Ostgaard}},
  \bibinfo{journal}{Nucl. Phys.} \textbf{\bibinfo{volume}{A437}},
  \bibinfo{pages}{487} (\bibinfo{year}{1985}).

\bibitem[{\citenamefont{Emery and Sessler}(1960)}]{ES60}
\bibinfo{author}{\bibfnamefont{V.~J.} \bibnamefont{Emery}} \bibnamefont{and}
  \bibinfo{author}{\bibfnamefont{A.~M.} \bibnamefont{Sessler}},
  \bibinfo{journal}{Phys. Rev.} \textbf{\bibinfo{volume}{114}},
  \bibinfo{pages}{1377} (\bibinfo{year}{1960}).

\bibitem[{\citenamefont{Gilbert and Cameron}(1965)}]{GC65}
\bibinfo{author}{\bibfnamefont{A.}~\bibnamefont{Gilbert}} \bibnamefont{and}
  \bibinfo{author}{\bibfnamefont{A.~G.~W.} \bibnamefont{Cameron}},
  \bibinfo{journal}{Can.\ J. Phys.} \textbf{\bibinfo{volume}{43}},
  \bibinfo{pages}{1446} (\bibinfo{year}{1965}).

\bibitem[{\citenamefont{Miller et~al.}(1989{\natexlab{a}})\citenamefont{Miller,
  Cole, and Quick}}]{MCQ89}
\bibinfo{author}{\bibfnamefont{H.~G.} \bibnamefont{Miller}},
  \bibinfo{author}{\bibfnamefont{B.~J.} \bibnamefont{Cole}}, \bibnamefont{and}
  \bibinfo{author}{\bibfnamefont{R.~M.} \bibnamefont{Quick}},
  \bibinfo{journal}{Phys.\ Rev.\ Lett.} \textbf{\bibinfo{volume}{63}},
  \bibinfo{pages}{1922} (\bibinfo{year}{1989}{\natexlab{a}}).

\bibitem[{\citenamefont{Levit and Alhassid}(1984)}]{LA84}
\bibinfo{author}{\bibfnamefont{S.}~\bibnamefont{Levit}} \bibnamefont{and}
  \bibinfo{author}{\bibfnamefont{Y.}~\bibnamefont{Alhassid}},
  \bibinfo{journal}{Nucl.\ Phys.\ A} \textbf{\bibinfo{volume}{413}},
  \bibinfo{pages}{439} (\bibinfo{year}{1984}).

\bibitem[{\citenamefont{Miller et~al.}(1989{\natexlab{b}})\citenamefont{Miller,
  Quick, and Cole}}]{MQC89}
\bibinfo{author}{\bibfnamefont{H.~G.} \bibnamefont{Miller}},
  \bibinfo{author}{\bibfnamefont{R.~M.} \bibnamefont{Quick}}, \bibnamefont{and}
  \bibinfo{author}{\bibfnamefont{B.~J.} \bibnamefont{Cole}},
  \bibinfo{journal}{Phys.\ Rev.\ C} \textbf{\bibinfo{volume}{39}},
  \bibinfo{pages}{1599} (\bibinfo{year}{1989}{\natexlab{b}}).

\bibitem[{\citenamefont{Dukelsky et~al.}(1991)\citenamefont{Dukelsky, Poves,
  and Retamosa}}]{DPR91}
\bibinfo{author}{\bibfnamefont{J.}~\bibnamefont{Dukelsky}},
  \bibinfo{author}{\bibfnamefont{A.}~\bibnamefont{Poves}}, \bibnamefont{and}
  \bibinfo{author}{\bibfnamefont{J.}~\bibnamefont{Retamosa}},
  \bibinfo{journal}{Phys. Rev. C} \textbf{\bibinfo{volume}{44}},
  \bibinfo{pages}{2872} (\bibinfo{year}{1991}).

\bibitem[{\citenamefont{Rossignoli et~al.}(1992)\citenamefont{Rossignoli,
  Quick, and Miller}}]{RQM92}
\bibinfo{author}{\bibfnamefont{R.}~\bibnamefont{Rossignoli}},
  \bibinfo{author}{\bibfnamefont{R.~M.} \bibnamefont{Quick}}, \bibnamefont{and}
  \bibinfo{author}{\bibfnamefont{H.~G.} \bibnamefont{Miller}},
  \bibinfo{journal}{Phys. Lett.} \textbf{\bibinfo{volume}{B 277}},
  \bibinfo{pages}{18} (\bibinfo{year}{1992}).

\bibitem[{\citenamefont{Civitarese and Schvellinger}(1994)}]{CS94}
\bibinfo{author}{\bibfnamefont{O.}~\bibnamefont{Civitarese}} \bibnamefont{and}
  \bibinfo{author}{\bibfnamefont{M.}~\bibnamefont{Schvellinger}},
  \bibinfo{journal}{Phys.\ Rev. C} \textbf{\bibinfo{volume}{49}},
  \bibinfo{pages}{1976} (\bibinfo{year}{1994}).

\bibitem[{\citenamefont{Yen and Miller}(1992)}]{YM92}
\bibinfo{author}{\bibfnamefont{G.~D.} \bibnamefont{Yen}} \bibnamefont{and}
  \bibinfo{author}{\bibfnamefont{H.~G.} \bibnamefont{Miller}},
  \bibinfo{journal}{Mod.\ Phys.\ Lett.\ A} \textbf{\bibinfo{volume}{7}},
  \bibinfo{pages}{1503} (\bibinfo{year}{1992}).

\bibitem[{\citenamefont{Yen and Miller}(1994)}]{YM94}
\bibinfo{author}{\bibfnamefont{G.~D.} \bibnamefont{Yen}} \bibnamefont{and}
  \bibinfo{author}{\bibfnamefont{H.~G.} \bibnamefont{Miller}},
  \bibinfo{journal}{Phys. Rev. C} \textbf{\bibinfo{volume}{50}},
  \bibinfo{pages}{807} (\bibinfo{year}{1994}).

\bibitem[{\citenamefont{Cole et~al.}(1998)\citenamefont{Cole, Miller, and
  Quick}}]{CMQ94}
\bibinfo{author}{\bibfnamefont{B.~J.} \bibnamefont{Cole}},
  \bibinfo{author}{\bibfnamefont{H.~G.} \bibnamefont{Miller}},
  \bibnamefont{and} \bibinfo{author}{\bibfnamefont{R.~M.} \bibnamefont{Quick}},
  \bibinfo{journal}{Mod. Phys Lett} \textbf{\bibinfo{volume}{A13}},
  \bibinfo{pages}{2705} (\bibinfo{year}{1998}).

\bibitem[{\citenamefont{Feng et~al.}(1979)\citenamefont{Feng, Gilmore, and
  Narducci}}]{FGN79}
\bibinfo{author}{\bibfnamefont{D.~H.} \bibnamefont{Feng}},
  \bibinfo{author}{\bibfnamefont{R.}~\bibnamefont{Gilmore}}, \bibnamefont{and}
  \bibinfo{author}{\bibfnamefont{L.~M.} \bibnamefont{Narducci}},
  \bibinfo{journal}{Phys.\ Rev.\ C} \textbf{\bibinfo{volume}{19}},
  \bibinfo{pages}{1119} (\bibinfo{year}{1979}).

\bibitem[{\citenamefont{Rossignoli and Plastino}(1985)}]{RP85}
\bibinfo{author}{\bibfnamefont{R.}~\bibnamefont{Rossignoli}} \bibnamefont{and}
  \bibinfo{author}{\bibfnamefont{A.}~\bibnamefont{Plastino}},
  \bibinfo{journal}{Phys.\ Rev.\ C} \textbf{\bibinfo{volume}{32}},
  \bibinfo{pages}{1041} (\bibinfo{year}{1985}).

\bibitem[{\citenamefont{Davis and Miller}(1987)}]{DM87}
\bibinfo{author}{\bibfnamefont{E.~D.} \bibnamefont{Davis}} \bibnamefont{and}
  \bibinfo{author}{\bibfnamefont{H.~G.} \bibnamefont{Miller}},
  \bibinfo{journal}{Phys.\ Lett.\ B} \textbf{\bibinfo{volume}{196}},
  \bibinfo{pages}{277} (\bibinfo{year}{1987}).

\bibitem[{\citenamefont{T.Duget and P.Bonche}(2004)}]{DB04}
\bibinfo{author}{\bibnamefont{T.Duget}} \bibnamefont{and}
  \bibinfo{author}{\bibnamefont{P.Bonche}},
  \bibinfo{howpublished}{nucl-th/0411023} (\bibinfo{year}{2004}).

\bibitem[{\citenamefont{Dean and Hjorth-Jensen}(2003)}]{DH03}
\bibinfo{author}{\bibfnamefont{D.~J.~.} \bibnamefont{Dean}} \bibnamefont{and}
  \bibinfo{author}{\bibfnamefont{M.}~\bibnamefont{Hjorth-Jensen}},
  \bibinfo{journal}{Rev. Mod.Phys.} \textbf{\bibinfo{volume}{75}},
  \bibinfo{pages}{607} (\bibinfo{year}{2003}).

\bibitem[{\citenamefont{Pathria}(1984)}]{P84}
\bibinfo{author}{\bibfnamefont{R.~K.} \bibnamefont{Pathria}},
  \emph{\bibinfo{title}{Statistical Mechanics}} (\bibinfo{publisher}{Pergamon
  Press}, \bibinfo{address}{Oxford}, \bibinfo{year}{1984}).

\bibitem[{\citenamefont{Solms et~al.}(1991)\citenamefont{Solms, Miller, and
  Plastino}}]{SMP91}
\bibinfo{author}{\bibfnamefont{F.}~\bibnamefont{Solms}},
  \bibinfo{author}{\bibfnamefont{H.~G.} \bibnamefont{Miller}},
  \bibnamefont{and} \bibinfo{author}{\bibfnamefont{A.}~\bibnamefont{Plastino}},
  \bibinfo{journal}{Phys.\ Lett.\ A} \textbf{\bibinfo{volume}{157}},
  \bibinfo{pages}{286} (\bibinfo{year}{1991}).

\end{thebibliography}

\end{document}